\documentclass[aps,prd,showpacs,preprintnumbers,nofootinbib,twocolumn]{revtex4}

\usepackage{bm}
\usepackage{latexsym}
\usepackage{dcolumn}
\usepackage{amsfonts,amssymb}
\usepackage{graphicx,epsfig}
\usepackage{psfrag}
\usepackage{amsmath,amssymb}

\def\beq{\begin{equation}}
\def\eeq{\end{equation}}
\def\br{\begin{eqnarray}}
\def\er{\end{eqnarray}}
\def\benu{\begin{enumerate}}
\def\eenu{\end{enumerate}}
\def\nn{\nonumber} 
\def\pa{{\partial}}
\def\l{\left}
\def\r{\right}    
\def\Hbar{\mathcal H}

\begin{document}

\preprint{hep-th/0403236}

\title{Trans-Planckian corrections to the primordial spectrum\\
in the infra-red {\it and}\/ the ultra-violet}
\author{S.~Shankaranarayanan}
\email[]{E-mail: shanki@ictp.trieste.it}
\affiliation{HEP Group, The Abdus Salam 
International Centre for Theoretical Physics,\\
Strada costiera 11, 34100 Trieste, Italy.}
\author{L.~Sriramkumar}
\email[]{E-mail: sriram@mri.ernet.in}
\affiliation{Harish-Chandra Research Institute, Chhatnag Road,\\
Jhunsi, Allahabad 211 019, India.}

\date{\today}


\begin{abstract}
Due to the tremendous red-shift that occurs during the inflationary
epoch in the early universe, it has been realized that trans-Planckian
physics may manifest itself at energies much lower than the Planck
energy.  The presence of a fundamental scale suggests that local
Lorentz invariance may be violated at sufficiently high energies.
Motivated by this possibility, recently, different models that violate
Lorentz invariance locally have been used to evaluate the
trans-Planckian corrections to the inflationary density perturbation
spectrum.  However, certain astrophysical observations seem to
indicate that local Lorentz invariance may be preserved to extremely
high energies.  In such a situation, to study the trans-Planckian
effects, it becomes imperative to consider models that {\it
preserve}\/ local Lorentz invariance {\it even as they contain a
fundamental scale}.\/ In this work, we construct one such model and
evaluate the resulting spectrum of density perturbations in the
power-law inflationary scenario.  While our model reproduces the
standard spectrum on small scales, it {\it naturally}\/ predicts a
suppression of power on large scales.  In fact, the spectrum we obtain
has some features which are similar to the one that has recently been
obtained from non-commutative inflation.  However, we find that the
amount of suppression predicted by our model is {\it far less}\/ than
that is required to fit the observations.  We comment on the fact
that, with a suitable choice of initial conditions, our approach can
lead to corrections at the infra-red {\it as well as}\/ at the
ultra-violet ends of the spectrum.
\end{abstract}
\maketitle


\section{Introduction and motivation} 

Inflation, a period of accelerated expansion in the high-energy phase
of the universe, is currently considered to be the best paradigm for
describing the early stages of the universe~\cite{txts,infltnmdls}.
The success of the inflationary paradigm rests on its ability to
explain not only the homogeneity of the background, but also the
characteristics of the inhomogeneities superimposed upon it.  The
inflationary epoch magnifies the tiny fluctuations in the quantum
fields present at the beginning of the epoch into classical
perturbations that leave an imprint as anisotropies in the cosmic
microwave background (CMB). These anisotropies in turn act as seeds
for the formation of the large-scale structure that we observe at the
present time as galaxies and clusters of galaxies.  With anisotropies
in the CMB being measured with higher and higher precision, we are
currently able to test the predictions of inflation better and better.

During the last couple of years or so, considerable amount of
attention has been devoted to examining the possibility that Planck
scale physics may leave a measurable imprint on the
CMB~\cite{mb01}--\cite{mb03}.  There are three reasons that have led
to such an enormous interest in the literature.  Firstly, there is no
unique model of inflation~\cite{infltnmdls}.  In many versions of
inflation, most notably in chaotic inflation, the period of inflation
lasts sufficiently long so that the comoving length scales that are of
cosmological interest today would have emerged from sub-Planckian
length scales at the beginning of inflation.  Hence, in principle,
quantum gravitational effects should have left their signatures on the
primordial spectrum of perturbations.  Secondly, in the simplest
models of inflation, the scale of the vacuum energy during the period
of exponential expansion is assumed to be $\sim 10^{16}\, {\rm GeV}$,
while the rate of the exponential expansion, viz. $H$, is considered
to be $\sim 10^{14}\, {\rm GeV}$.  These enormous energies suggest
that during the inflationary epoch, various (as yet, unknown?)
high-energy processes could have been activated which would have left
their signatures on the primordial perturbation spectrum.  These
signatures on the perturbation spectrum in turn will leave their
imprints on the CMB.  Recent measurements of the CMB anisotropies by,
say, WMAP~\cite{MAP}, strongly indicate a primordial spectrum that is
nearly scale-invariant, just as the inflationary scenario predicts.
Therefore, further precise and accurate measurements of the CMB
anisotropies by experiments such as PLANCK~\cite{PLANCK} can, in
principle, provide us with the form of the corrections to the
scale-invariant primordial perturbation spectrum.  Lastly and, more
importantly, the first results of the temperature-temperature
correlation spectrum of the WMAP data show that the power in the
quadrupole and (to a lesser extent) the octopole moment of the CMB are
{\it lower than}\/ as expected in the best fitting cold dark matter
models~\cite{MAP}.  The deficit of power in the quadrapole moment of
the CMB power spectrum can not be explained within the context of the
standard inflationary models (unless these models are fine-tuned
\cite{contaldi}) and suggests a possible signature of the Planck scale
physics~\cite{nci}.

Several groups have pointed out to these possibilities and, broadly,
there have been two approaches in the literature in order to study
these effects. In the first approach, which is now commonly referred
to as the minimal trans-Planckian approach, the specific nature of
trans-Planckian physics is not assumed, but is rather described by the
boundary conditions imposed on the mode at the cut-off
scale~\cite{minimaltp,act03}. In the second approach, one
incorporates quantum gravitational effects by introducing the
fundamental length scale into the standard field theory in a
particular fashion and, it should be emphasized here that, the
resulting modified theory may not even preserve local Lorentz
invariance.  Indeed, most of the attempts in this direction have
involved models which {\it break}\/ local Lorentz invariance.  Such
models include those which introduce non-linear dispersion
relations~\cite{mb01,niemeyer,earliertp,br,shanki03} as well as the
approaches which utilize the generalized uncertainty
principle~\cite{guptp} and non-commutative geometry~\cite{nctp,nci}.
(For a recent review on these various approaches, see
Ref.~\cite{mb03}.)

However, theoretically, there exists no apriori reason to believe that
Lorentz invariance may be broken at the scales of inflation. More
importantly, recent observations of photons inferred to arise from
synchrotron emission off electrons in the Crab nebula have placed
bounds on dispersive corrections involving additional powers of the
particle momentum to be at or above the Planck scale~\cite{liv}. In
such a situation, in order to study the trans-Planckian effects on the
primordial perturbation spectrum, it becomes important that we also
consider models which preserve Lorentz invariance {\it even as they
contain a fundamental scale}.\/ In this work, we consider one such
model and evaluate the resulting spectrum of density perturbations
during inflation in this model.

The model we shall consider is as follows: We shall {\it assume}\/
that, due to the Planck-scale effects, the standard $k-$space
propagator for a massless scalar field in the Minkowski vacuum is
modified in a particular manner.  We shall introduce the high-energy
scale $k_{\rm c}$ into the $k-$space propagator in a Lorentz invariant
manner and, as we shall point out later, the modification we consider
can be said to be minimal in nature.  We find that the resulting
modified Wightman function can be expressed as the difference of the
Wightman functions of the massless field and a massive field of mass
$k_{\rm c}$.  We shall further assume that the form of the modified
Wightman function in the $x-$space remains the same in an inflationary
background as well, i.e. it is the difference of the massless and the
massive Wightman functions.  We then use this modified Wightman
function to evaluate the corrections to the spectrum of density
perturbations for the power-law inflationary scenario.  We show that
our model naturally predicts a suppression of power at the large
scales.  We find that the modified spectrum we obtain has some
resemblance to the spectrum that has been obtained in the
non-commutative inflationary scenario~\cite{nci}.  However, the amount
of suppression predicted by our model turns out to be {\it far less}\/ 
than that seems to be required to fit the WMAP data.

The rest of the paper is organized as follows.  In Section~(II), we
outline the model we shall consider, present the motivations behind
the model and also point out its attractive features.  In
Section~(III), we shall briefly review the spectrum of fluctuations in
the standard power-law inflationary scenario.  In Section~(IV), we
shall evaluate the corrections to the spectrum of fluctuations using
our model for the power-law inflation.  Finally, in Section~(V), we
shall summarize the results of our analysis and also discuss its
implications.

Before we proceed, a few comments on the notations we shall use are in
order.  The metric signature we shall adopt is $(+, -,-, -)$, we shall
set $\hbar=c=1$ and we shall denote the set of four coordinates
$x^{\mu}$ simply as ${\tilde x}$.  Also, the quantum field $\Psi$ we
shall consider will be a minimally coupled scalar field.

\section{The model}\label{sec:mdl}

In the inflationary scenario, the primordial perturbation spectrum is
given by the Fourier transform of the Wightman function of a
quantized, massless scalar field in the inflating
background~\cite{txts,infltnmdls}. Therefore, in order to understand
the effects of Planck-scale physics on the perturbation spectrum, we
need to understand as to how quantum gravitational effects will modify
the propagator of a scalar field in an inflationary background.
However, due to the lack of a clear understanding of the Planck-scale
effects, often, one is forced to consider models constructed by
hand---models which are supposed to be {\it effective theories}\/
obtained by integrating out the gravitational degrees of freedom.
 
For reasons we had outlined in the introductory section, in this work,
we would like to consider a model that is locally Lorentz invariant
and also contains the high-energy scale. Recall that, in the
Minkowski vacuum in flat space-time, the propagator for a massless
scalar field in $k-$space is given by
\beq
G_{0}^{+}(k) = \l(\frac{1}{{\bar k}^2}\r),\label{eq:Gfnk}
\eeq
where ${\bar k}^2\equiv (k^{\mu}k_{\mu})$. Planck-scale effects are
expected to modify the propagator even in flat space-time (see, for
e.g., Refs.~\cite{smrdGfn} and references therein). If we require
that the effective theory describing the modified propagator be
Lorentz invariant, then, in flat space-time, the modified propagator
can only be a function of ${\bar k}$. We shall {\it assume}\/ that
the standard propagator (\ref{eq:Gfnk}) for the massless field is
modified to
\beq
G_{\rm M}^+(k) = \l(\frac{1}{{\bar k}^2\, [1- \alpha^2 ({\bar
k}/k_{\rm c})^2]}\r),
\label{eq:mGfnk1}
\eeq
where $\alpha^2$ is an arbitrary positive constant fixed by the
complete theory and $k_{\rm c}$ denotes the cut-off scale which we
shall assume to be, say, three to five orders of magnitude above the
Hubble scale during inflation. Such a modification can be considered
to be minimal as it contains only terms of order ${\bar k}^4$ and does
not contain any higher order terms. (We introduce the ${\bar k}^4$
term rather than the immediately higher order ${\bar k}^3$ term, as it
makes the calculation of the resulting perturbation spectrum far more
tractable.) Note that the modified propagator $G_{\rm M}^{+}(k)$ can
be expressed as
\beq
G_{\rm M}^{+}(k)= \l(\frac{1}{{\bar k}^2}\right) 
-\l(\frac{1}{{\bar k}^2 - k_{\rm c}^2}\r),\label{eq:mGfnk2}
\eeq
where we have absorbed $\alpha$ into $k_{\rm c}$. Then, in $x-$space,
the Wightman function corresponding to the above $k-$space propagator
is given by
\beq
G_{\rm M}^{+}\l({\tilde x},{\tilde x'}\r) 
= G_{0}^{+}\l({\tilde x},{\tilde x'}\r) 
- G_{k_{\rm c}}^{+}\l({\tilde x},{\tilde x'}\r),\label{eq:mGfnx}
\eeq
where $G_{0}^{+}\l({\tilde x},{\tilde x'}\r)$ and $G_{k_{\rm
c}}^{+}\l({\tilde x},{\tilde x'}\r)$ denote the the Wightman functions
of the massless field and a massive field of mass $k_{\rm c}$. In
order to evaluate the resulting modification to the primordial
perturbation spectrum, we shall assume that the above modified
Wightman function retains the same form (i.e. it is the difference of
the Wightman functions of massless and massive fields) in the
inflationary background as well.

Apart from the fact that we required a minimal (and tractable) Lorentz
invariant model, the other motivation for the above model is as
follows: The divergence structure of quantum field theory is expected
to vastly improve when the quantum gravitational effects have been
taken into account. In particular, propagators are expected to be
finite in the limit when the two space-time points
coincide~\cite{smrdGfn}. Though, the modified Wightman function
$G_{\rm M}^{+}({\tilde x}, {\tilde x'})$ we have constructed will not
actually be finite as ${\tilde x}\to {\tilde x'}$, it certainly turns
out to be far less divergent than the original Wightman function. (In
this sense, our approach can be said to be motivated by the
Pauli-Villars regularization procedure.) For instance, in the
Minkowski vacuum, while the original Wightman function will diverge as
$0^{-2}$ in the coincident limit, the modified function
(\ref{eq:mGfnx}) will diverge as ${\rm ln}\, 0$ (see, for instance,
Ref.~\cite{bc83}). Moreover, we should add that a similar approach
has been considered earlier to study the effects of trans-Planckian
physics on Hawking radiation from black holes~\cite{hb96}.

At this stage of our discussion, it is important that we compare our
model with the models that introduce non-linear dispersion relations
to study the trans-Planckian effects on the primordial perturbation
spectrum~\cite{mb01,niemeyer,earliertp,br,shanki03}. In these models,
there has been the growing issue of the choice of the vacuum state and
the back-reaction of the trans-Planckian modes on the inflationary
background.  It has been shown that the back-reaction effects during
inflation become important for those dispersion relations which induce
large corrections to the scale-invariant spectrum~\cite{br}. This
feature essentially arises due to the fact that adiabatic evolution
can occur only if the dispersion relations either grow or, at least,
reach a plateau for large values of the
wavenumber~\cite{niemeyer,earliertp}. In contrast, in our model, the
modified propagator and the resulting modification to the perturbation
spectrum involve modes of massless and massive fields which always
evolve adiabatically.

Before proceeding to the technical aspects, it is necessary that we
clarify certain conceptual issues related to our model. In the
conventional models of quantum field theory (and also in the other
models that incorporate quantum gravitational corrections such as the
dispersive field theory models), the Lagrangian is a function of the
fields and their first derivatives in time. However, our model
contains second order time derivatives in the Lagrangian.  It can be
easily shown that, in flat space-time, the Feynman propagator
corresponding to the modified Wightman function~(\ref{eq:mGfnx})
satisfies the equation (see, for e.g., Ref.~\cite{bc83}; in this
context, also see, Refs.~\cite{hb96,hh02})
\beq
\l[\Box + \l(\frac{1}{k_{\rm c}^2}\r) \Box^2 \r] 
G^F_{\rm M}({\tilde x},{\tilde x}')  
= -\, \delta^{(4)}\l({\tilde x} -{\tilde x}'\r)
\eeq
which in turn can be obtained from the following action 
\beq
S[\Psi] = \int d^4x\, \l[\l(\frac{1}{2}\r) 
\pa_{\mu}\Psi\, \pa^{\mu}\Psi - 
\l(\frac{1}{2\, k_{\rm c}^2}\r) (\Box \Psi)^2 \r],
\label{eq:EOM-scalar}
\eeq
where $\Psi$ denotes a scalar field and $k_{\rm c}$ the cut-off
scale. 

Theories described by higher derivative Lagrangians are well-known to
have unsatisfactory properties~\cite{simon90,hh02}. They are known to
contain additional degrees of freedom, the energy is not bounded from
below and, also, the solutions to the equations of motion are not
uniquely determined by the initial values of the fields and their
first time derivatives. Clearly, these features can be considered to
be unwelcome when we are dealing with Lagrangians that are expected to
parameterize {\it small}\/ deviations from a well-understood theory.

In such a situation, we shall adopt the following point-of-view
concerning our model. As mentioned earlier, we consider our model to
be an effective theory of the inflaton field obtained by integrating
out the Planck scale effects. Non-locality often arises in such
effective theories. Effective theories in which the non-locality can
be regulated by a small parameter are known to have perturbation
expansions with higher derivatives~\cite{simon90}. We believe that
our model is the leading term in the perturbation expansion in the
small parameter $(1/k_{\rm c})$ of a non-local effective theory.
Moreover, we expect that, in the complete theory, we may not encounter
the difficulties that we mentioned in the previous paragraph.

\section{Spectrum of perturbations from inflation---the standard
result}

In this section, we shall briefly rederive the standard spectrum 
of perturbations in the power-law inflationary scenario.

Consider a flat Friedmann universe described by the line-elements
\beq
ds^2=\l(dt^2-a^{2}(t)\, d{\bf x}^2\r) =a^{2}(\eta)\l(d\eta^{2} 
- d{\bf x}^2\r),\label{eq:frw}
\eeq 
where $t$ is the cosmic time, $a(t)$ is the scale factor and $\eta=\int 
\l[dt/a(t)\r]$ denotes the conformal time. 
Power-law inflation corresponds to the situation where the scale 
factor $a(t)$ is given by
\beq 
a(t) = \l(a_0 \, t^p\r),
\eeq
where $p > 1$. 
In terms of the conformal time $\eta$, this scale factor can be 
written as
\beq
a(\eta) = \l(-\Hbar \, \eta \r)^{(\beta+1)}, 
\eeq
where $\beta$ and $\Hbar$ are given by
\beq
\beta = -\l(\frac{2p-1}{p - 1}\r)
\quad{\rm and}\quad
\Hbar = \l[(p - 1) \, a_0^{1/p}\r].
\eeq
Note that $\beta \le -2$ and $\Hbar$ denotes the characteristic energy
scale associated with inflation. Also, the case of exponential
expansion corresponds to $\beta=-2$ with $\Hbar$ actually being equal
to the Hubble scale.

Metric fluctuations (both scalar as well as the tensor perturbations)
during the inflationary epoch can be modeled by a massless, minimally
coupled scalar field, say, $\Psi$~\cite{txts,perturbations}.  The
field $\Psi$ propagating in a background described by the
line-elements~(\ref{eq:frw}) satisfies the following Klein-Gordon
equation:
\beq 
\Box\Psi 
\equiv \l(\frac{\pa^2\Psi}{\pa\eta^2}\r) + \l(\frac{2}{a}\r)
\l(\frac{da}{d \eta}\r) \l(\frac{\pa \Psi}{\pa \eta}\r)
- \nabla^2\Psi = 0.
\eeq
The symmetry of the Friedmann metric allows us to decompose the 
normal modes of the scalar field as 
\beq
\psi_{\bf k}({\tilde x})
= \l(\frac{1}{(2\pi)^{3/2}}\r)
\l(\frac{\mu_{k}(\eta)}{a(\eta)}\r)\, e^{i{\bf k}\cdot{\bf x}},
\label{eq:nrmlmds}
\eeq
where ${\bf k}$ is the comoving wave vector and the function 
$\mu_{k}$ satisfies the differential equation
\beq
\mu_{k}''+\l[k^2 - \l(\frac{a''}{a}\r)\r]\mu_{k}
=0\label{eq:demu0}
\eeq
with the primes denoting differentiation with respect to $\eta$ 
and $k=|{\bf k}|$.
 
On quantization, the scalar field can be expressed in terms of the 
normal modes $\psi_{\bf k}({\tilde x})$ as follows:
\beq {\hat \Psi}\l(\eta, {\bf x}\r)
=\int d^{3}{\bf k}\, 
\l[{\hat a}_{\bf k}\, \psi_{\bf k}({\tilde x}) 
+ {\hat a}_{\bf k}^{\dag}\, \psi_{\bf k}^{*}({\tilde x})\r],
\label{eq:Psidcmpstn}  
\eeq 
where the creation and the annihilation operators ${\hat a}_{\bf k}$
and ${\hat a}_{\bf k}^{\dag}$ obey the usual commutation relations.
As the perturbations are assumed to be induced by the fluctuations in
the free quantum field ${\hat\Psi}$, the power spectrum as well as the
statistical properties of the perturbations are entirely characterized
by the Wightman functions of the quantum field.  Therefore, the power
spectrum of the perturbations per logarithmic interval,
viz. $\left[k^3\, {\cal P}_{\Psi}(k)\right]$, is given
by~\cite{txts,infltnmdls}
\beq
\int\limits_{0}^{\infty} \l(\frac{dk}{k}\r)\, 
\left[k^3\; {\cal P}_{\Psi}(k)\right] 
=\langle 0 \vert {\hat \Psi}^2 (\eta, {\bf x})\vert 0\rangle 
= G^{+}_{0}\l({\tilde x},{\tilde x}\r),\label{eq:psdfntn}
\eeq 
where $\vert 0\rangle$ is the vacuum state (defined as $\hat a_{\bf k}
\vert 0\rangle=0$ $\forall\; {\bf k}$), $G^{+}_{0}({\tilde x}, {\tilde 
x'})$ denotes the Wightman function of the massless quantum field 
${\hat \Psi}$ and the spectrum is to be evaluated when the modes
leave the Hubble radius. 
Using the decomposition (\ref{eq:Psidcmpstn}), the perturbation 
spectrum per logarithmic interval can then be written in terms of 
the modes $\mu_{k}$ as 
\beq 
\left[k^3\; {\cal P}_{\Psi}(k)\right]
=\left(\frac{k^3}{2\pi^2}\right)\, \left(\frac{\vert
\mu_{k}\vert}{a}\right)^2\label{eq:ps}
\eeq 
and the expression on the right hand side is to be evaluated when the
physical wavelength $(k/a)^{-1}$ of the mode corresponding to the
wavenumber ${\bf k}$ equals the Hubble radius $H^{-1}$, where
$H=(a'/a^2)$.  In the remainder of this section, we shall evaluate the
spectrum of perturbations in power-law inflation\footnote{In this
work, we do not distinguish between the scalar and the tensor
perturbations since, in power-law inflation, the power spectrum of
these perturbations are related by a constant
factor~\cite{txts,mb01,mb03}.}.

For the case of power-law inflation, the exact solution to the
differential equation~(\ref{eq:demu0}) is well-known.  Nevertheless,
due to its utility in the next section (where we need to evaluate the
modes of a massive field in the WKB approximation), we shall rewrite
the differential equation (\ref{eq:demu0}) in terms of a new set of
independent and dependent variables $(x,u_{k})$ which are related to
the old set $(\eta,\mu_{k})$ by the relations~\cite{wms97,ms03}
\beq
x = \ln \l(\frac{\beta+1}{k\eta}\r)\label{eq:xdfntn}
\eeq
and
\beq
\mu_{_{k}}= {\rm e}^{-(x/2)}\, u_{k}.\label{eq:udfntn}
\eeq
In terms of the new variables, the differential 
equation (\ref{eq:demu0}) can be written as
\beq
\l(\frac{d^2 u_{k}}{dx^2}\r) + \l[(\beta+1)^2\, {\rm e}^{-(2 x)} - 
\l( \beta + \frac{1}{2} \r)^2\r]u_{k} = 0\label{eq:demu0ux}
\eeq
and the general solution to this differential equation is given 
by (see, for e.g., Ref.~\cite{as64}, p.~362)
\br
\!\!\!\!\!\!\!\!\!\!\!\!
u_{k}(x) & = & \biggl(A(k)\, H_{-\l(\beta+\frac{1}{2}\r)}^{(2)}
\l[(\beta + 1)\, {\rm e}^{-x} \r]\nn \\
& &\qquad\quad+\,   
B(k) \, H_{-\l(\beta + \frac{1}{2}\r)}^{(1)}
\l[(\beta + 1)\, {\rm e}^{-x}\r]
\biggl).\label{eq:gsmu0}
\er
The quantities $H_{\nu}^{(1)}$ and $H_{\nu}^{(2)}$ in the above
solution are the Hankel functions of the first and the second kind (of
order $\nu$), respectively, and the $k$-dependent constants $A(k)$ and
$B(k)$ are to be fixed by choosing suitable initial conditions for the
modes at the beginning of inflation.

In the standard inflationary cosmology, the initial conditions are
imposed on sub-Hubble scales, i.e. when the physical wavelengths
$(k/a)^{-1}$ of the modes are much smaller than the Hubble radius
$H^{-1}$.  In this limit, the modes do not feel the curvature of the
space-time and, hence, in terms of $\eta$, reduce to $e^{\pm ik\eta}$.
The assumption that the quantum field is in the vacuum state then
requires that $\mu_{k}(\eta)$ is a positive frequency mode at
sub-Hubble scales, i.e. it has the asymptotic form (see, for instance,
Ref.~\cite{ms03})
\beq
\lim _{\l(k/aH\r)\to \infty}\mu_{k}(\eta)
\to \l(\frac{1}{\sqrt{2k}}\r)\, {\rm e}^{-ik\eta}.\label{eq:stic}
\eeq
In terms of the variable $x$, sub-Hubble scales (i.e. $(k/a) \gg H$)
correspond to the limit $x \to - \infty$, super-Hubble scales (i.e.
$(k/a)\ll H$) correspond to $x \to \infty$ and Hubble exit of the
modes occur at $x=0$.  Therefore, in terms of $x$ and $u_{k}$, the
condition (\ref{eq:stic}) reduces to
\beq
\lim _{x \rightarrow -\infty} u_{k}(x) \to  
\l(\frac{1}{\sqrt{2k}}\r)\, {\rm e}^{(x/2)}\, 
\exp{-i[(\beta + 1)\, {\rm e}^{-x}]}.\label{eq:sticux}
\eeq
This can be achieved by setting $B(k)$ to zero and choosing $A(k)$
to be
\beq
A(k) = \l(\frac{\pi\, (\beta + 1)}{4k}\r)^{1/2}\, 
{\rm e}^{i \pi \beta/2}
\eeq
in Eq.~(\ref{eq:gsmu0}), so that we have\footnote{Using the asymptotic 
behavior of the Hankel function, viz. (cf.~Ref.~\cite{as64}, p.~364)
\beq
\lim_{z\to \infty} H^{(2)}_{\nu}(z)\longrightarrow
\l({\frac{2}{\pi z}}\r)^{1/2}\, {\rm e}^{-i\l[z-(\pi \nu /2)-(\pi /4)\r]},
\eeq
it is straightforward to check that the function $u_{k}(x)$ in 
Eq.~(\ref{eq:ux}) has indeed the required limit~(\ref{eq:sticux}).}
\beq
u_{k}(x) = \l(\frac{\pi\, (\beta + 1)}{4k}\r)^{1/2}\, 
{\rm e}^{i(\pi \beta/2)}\,
H^{(2)}_{-\l(\beta + \frac{1}{2}\r)}\l[(\beta + 1)\, 
{\rm e}^{-x}\r]\label{eq:ux}
\eeq
and it should be mentioned here that the vacuum state associated with
this mode is often referred to in the literature as the Bunch-Davies
vacuum~\cite{bd78}.  The spectrum of perturbations can now be obtained
by substituting the mode (\ref{eq:ux}) in the expression~(\ref{eq:ps})
[with $(\eta, {\mu_k})$ related to $(x,u_{k})$ by
Eqs.~(\ref{eq:xdfntn}) and~(\ref{eq:udfntn})] and finally setting
$x=0$.  We obtain the power spectrum, at Hubble exit, to
be~\cite{plps}
\beq 
\left[k^3\; {\cal P}_{\Psi}(k)\right] 
= C \, \l(\frac{\Hbar^2}{2 \pi ^2}\r) \, 
\l(\frac{k}{\Hbar}\r)^{2(\beta+2)} \, , \label{eq:psplfv}
\eeq
where $C$ is given by 
\beq
C = \l(\frac{\pi}{4}\r)\, (1 + \beta)^{-2 (\beta + 1)}\, 
\biggl \vert H_{-\l(\beta + \frac{1}{2}\r)}^{(2)}(\beta+1)\biggr\vert^{2}.
\eeq
Note that, in obtaining the above expression, we have evaluated the
power spectrum when the modes leave the Hubble radius.  In the
literature (see, for instance, Ref.~\cite{plps}), the spectrum of
perturbations is evaluated at the super-Hubble scales.  These two
power spectra typically differ in their amplitude by a numerical
factor of order unity.

\section{Corrections to the standard perturbation spectrum}
\label{sec:crrctns}

As we had discussed in Section~(\ref{sec:mdl}), in our model, the
modified Wightman function of a massless field $G_{\rm M}\l({\tilde
x}, {\tilde x'}\r)$ can be expressed as the difference of the Wightman
functions of the massless field and a massive field of mass $k_{\rm
c}$, viz. $G_{0}^{+}\l({\tilde x},{\tilde x'}\r)$ and $G_{k_{\rm
c}}^{+}\l({\tilde x},{\tilde x'}\r)$ [cf.~Eq.~(\ref{eq:mGfnx})].
Therefore, following Eq.~(\ref{eq:psdfntn}), we can define the
resulting modified perturbation spectrum per logarithmic interval,
viz. $\left[k^3\; {\cal P}_{\Psi}(k)\right]_{\rm M}$, as follows:
\br 
\!\!\!\!\!\!\!\!\!\!
\int\limits_{0}^{\infty} 
\l(\frac{dk}{k}\r)\, \left[k^3\; {\cal P}_{\Psi}(k)\right]_{\rm M}
&=& G_{\rm M}^{+}\l({\tilde x},{\tilde x}\r)\nn\\
&=& G_{0}^{+}\l({\tilde x},{\tilde x}\r) 
- G_{k_{\rm c}}^{+}\l({\tilde x},{\tilde x}\r).\label{eq:mpsdfntn}
\er
The massive field can be decomposed and quantized in terms of the
normal modes along the same lines as we had done in
Eqs.~(\ref{eq:nrmlmds}) and (\ref{eq:Psidcmpstn}) for the case of the
massless field.  If we now assume that the massive field is also in
the Bunch-Davies vacuum, then the modified perturbation spectrum can
be written as
\beq 
\left[k^3\; {\cal P}_{\Psi}(k)\right]_{\rm M} 
=\l(\frac{k^3}{2\pi^2}\r)\, 
\l[\l(\frac{\vert \mu_{k}\vert}{a}\r)^2 
- \l(\frac{\vert{\bar \mu}_{k}\vert}{a}\r)^2\r],
\label{eq:psm}
\eeq 
where, as before, $\mu_{k}$ denotes the modes of the massless field
and ${\bar \mu}_{k}$ denotes the modes of the scalar field with mass
$k_{\rm c}$ which satisfies the differential equation
\beq
{\bar \mu}_{k}''+\l[k^2+(k_{\rm c}\, a)^2
- \l(\frac{a''}{a}\r)\r]{\bar \mu}_{k} =0.\label{eq:demukc}
\eeq
Moreover, as in the standard case, the expression on the right hand
side of Eq.~(\ref{eq:psm}) is to be evaluated at the time when the
modes of the massless and the massive fields leave the Hubble radius.
On comparing the form of the original perturbation
spectrum~(\ref{eq:ps}) with the modified spectrum~(\ref{eq:psm}), it
is clear that the corrections to the standard spectrum arise as a
result of the contribution due to the massive modes.

Before we proceed further with the evaluation of the corrections, we
would like to stress the following point: In the standard inflationary
scenario, it is well-known that the amplitude of the spectrum
corresponding to the massive modes decays at the super-Hubble scales.
(This can be easily shown for the case of exponential expansion,
wherein the solutions to the massive modes are exactly known---in this
context, see, for e.g., Ref.~\cite{lv04}).  As we had pointed out in
the previous paragraph, in our model, the trans-Planckian corrections
to the standard adiabatic primordial spectrum arise due to the massive
modes.  Hence, within the standard inflationary picture, the amplitude
of these corrections would be expected to decay at the super-Hubble
scales.  However, as the massive modes we have considered are supposed
to represent the trans-Planckian corrections to the standard, massless
modes, we shall assume that the mechanism that `freezes' the amplitude
of the standard adiabatic spectrum at super Hubble scales will also
`freeze' the amplitude of the trans-Planckian corrections at their
value at Hubble exit.  Hence, in what follows, we shall evaluate the
corrections to the standard power spectrum when the massive modes
leave the Hubble radius.

We can now introduce a new set of variables $(x,{\bar u}_{k})$ which
are related to the old set $(\eta,{\bar \mu}_{k})$ exactly as in the
massless case through the relations~(\ref{eq:xdfntn})
and~({\ref{eq:udfntn}).  In terms of the new variables, the
differential equation~(\ref{eq:demukc}) can be written as
\beq
\l(\frac{d^2 {\bar u}_{k}}{dx^2}\r) + \omega^2(x)\, {\bar u}_{k}(x) 
= 0,\label{eq:demukcux} 
\eeq
where $\omega^2(x)$ is given by 
\br
\!\!\!\!\!\!\!\!\!\!
\omega^2(x) 
\!&=&\! \Biggl[(\beta + 1)^2\, e^{-2x} 
- \l(\beta + \frac{1}{2}\r)^2\nn\\
& &\!+\, \l(\frac{k_c}{\Hbar}\r)^2\, 
\l(\frac{\Hbar(\beta + 1)}{k}\r)^{2(\beta + 2)}\, 
{\rm e}^{-2(\beta+2)x}\Biggr].\;\label{eq:omega2}
\er
Unlike the massless case, the exact solution to the differential
equation~(\ref{eq:demukcux}) for power-law inflation is not known.
(The solution is known only for the special case of $\beta=-2$ which
corresponds to exponential expansion.)  Hence, we shall obtain the
solution in the WKB approximation.  As we shall show, for a massive
field such that $k_{\rm c}\gg \Hbar$, the WKB approximation turns out
to be valid for all $x$ over a range of values of $\beta$ and $k$ of
our interest.  For our discussion below, we shall assume that $10^{-5}
\lesssim \l(\Hbar/k_{\rm c}\r) \lesssim 10^{-3}$. 

To begin with, note that, for $\beta \le 2$ and $k_{\rm c}\gg \Hbar$, 
$\omega^2(x)$ remains positive for all values of $x$.
Therefore, the WKB solutions to the differential 
equation~(\ref{eq:demukcux}) are given by (see, for e.g.,
Ref.~\cite{ms03})
\beq
{\bar u}^{\rm WKB}_{k}(x) = \l(\frac{1}{\sqrt{\omega(x)}}\r)\,
\exp\pm\, i\int^{x}dx'\, \omega(x'),\label{eq:wkb}
\eeq
where ${\bar u}^{\rm WKB}_{k}$ satisfies the differential equation
\beq
\l(\frac{d^2 {\bar u}^{\rm WKB}_{k}}{dx^2}\r) 
+ \biggl[\omega ^2(x) - Q(x)\biggr] {\bar u}^{\rm WKB}_{k}= 0
\label{eq:dewkb}
\eeq
with the quantity $Q(x)$ defined as
\begin{equation}
Q(x) 
= \l[\l(\frac{3}{4 \omega ^2}\r)\l(\frac{d \omega}{dx}\r)^2
-\l(\frac{1}{2\omega}\r) \l(\frac{d^2 \omega}{d x^2}\r)\r].
\end{equation}
It is then clear from Eq.~(\ref{eq:dewkb}) that the WKB 
solution~(\ref{eq:wkb}) will be a good approximation to mode 
function $\bar{u}(x)$, only if the following condition is 
satisfied~\cite{ms03}:
\beq
\biggl \vert \l(\frac{Q}{\omega^2}\r)\biggr \vert \ll 1.
\label{eq:wkbcndtn}
\eeq

For $\omega^{2}$ given by Eq.~(\ref{eq:omega2}), we find that
the quantity $(Q/\omega^2)$ can be written as
\begin{widetext}
\br
\l(\frac{Q}{\omega^2}\r) 
&=& \Biggl\{\l(\frac{5}{4\, \omega^6}\r) \, 
\l[(\beta+1)^2\, {\rm e}^{-2x}
+(\beta + 2) 
\l(\frac{k_c^2}{\Hbar^2}\r)
\l(\frac{\Hbar\, (\beta + 1)\, 
{\rm e}^{-x}}{k}\r)^{2(\beta + 2)}\r]^2\nn\\
& &\qquad\qquad\qquad\qquad\qquad\quad
-\,\l(\frac{1}{\omega^4}\r)
\l[(\beta+1)^2\, {\rm e}^{-2x} 
+(\beta + 2)^2 \l(\frac{k_c^2}{\Hbar^2}\r)
\l(\frac{\Hbar\, (\beta + 1)\, 
{\rm e}^{-x}}{k}\r)^{2(\beta + 2)}\r]\Biggr\}.
\label{eq:Qomega2}
\er
\end{widetext}
Our task now is to examine whether this $(Q/\omega^2)$ indeed satisfies 
the condition (\ref{eq:wkbcndtn}) in all the three regimes, viz. on the 
sub-Hubble and the super-Hubble scales as well as at Hubble exit.
Let us first consider the case of exponential expansion, i.e. when 
$\beta=-2$.
In such a case, $\omega^2$ and $(Q/\omega^2)$ simplify to 
\beq
\omega^2 
= \l[{\rm e}^{-2x} + \l(k_c/\Hbar\r)^2\, 
- \l(9/4\r)\r]\label{eq:omega2ei}
\eeq
and
\beq
\l(\frac{Q}{\omega^2}\r) 
= \l[\l(\frac{5}{4\, \omega^6}\r)\, {\rm e}^{-4x}
-\l(\frac{1}{\omega^4}\r)\, {\rm e}^{-2x}\r]. 
\label{eq:Qomega2ei}
\eeq
We find that this expression reduces to
\beq
\lim_{x\to -\infty}\;
\biggl \vert \l(\frac{Q}{\omega^2}\r)\biggr\vert 
\simeq \l({\rm e}^{2 x}/4\r) \to 0
\eeq
on the sub-Hubble scales, to
\beq
\lim_{x\to\infty}\;
\biggl \vert\l(\frac{Q}{\omega^2}\r)\biggr \vert
\simeq \l(\Hbar/k_{c}\r)^4\, {\rm e}^{-2x}\to 0
\eeq
on the super-Hubble scales and to 
\beq
\lim_{x\to 0}\;
\biggl \vert \l(\frac{Q}{\omega^2}\r)\biggr\vert 
\simeq \l(\frac{\Hbar}{k_{c}}\r)^4\label{eq:WKBveiHe}
\eeq
at Hubble exit.
Since $\l(\Hbar/k_{c}\r)\ll 1$, clearly, $(Q/\omega^2)\ll 1$ at
Hubble exit and, therefore, the WKB approximation for the massive
modes is valid for all $x$ and $k$.

Let us now consider the case of power-law inflation, i.e. when 
$\beta<-2$.
We find that, the expression (\ref{eq:Qomega2}) for $(Q/\omega^2)$ 
reduces to
\beq
\lim_{x\to -\infty}\;
\biggl \vert \l(\frac{Q}{\omega^2}\r)\biggr\vert 
\simeq \l[\frac{e^{2 x}}{4\, (\beta + 1)^{2}}\r] \to 0
\eeq
on the sub-Hubble scales, to
\br
\!\!\!\!\!\!\!\!
\lim_{x\to\infty}\;
\biggl \vert\l(\frac{Q}{\omega^2}\r)\biggr \vert
&\simeq&
\l(\frac{\Hbar^2}{4\, k_{c}^2}\r)\,
\l(\frac{k}{\Hbar\,(\beta+1)}\r)^{2(\beta+2)}\nn\\
& &\qquad\;\times\; (\beta+2)^2\, e^{2(\beta + 2) x}  \to 0
\er
on the super-Hubble scales and to
\beq
\lim_{x\to 0}\;
\biggl \vert \l(\frac{Q}{\omega^2}\r)\biggr \vert \simeq
\l(\frac{\Hbar^2}{4\, k_{c}^2}\r)\,
\l(\frac{k}{\Hbar\, (\beta+1)}\r)^{2(\beta+2)}\, (\beta+2)^2
\eeq
at Hubble exit.
Evidently, the WKB approximation is valid at the sub-Hubble and the
super-Hubble scales.
However, at Hubble exit, the validity of the approximation depends
on the values of $\beta$ and $k$ and, for a given value of $\beta$,
the approximation breaks down at a sufficiently small value of $k$.
Fig.~\ref{fig:wkb} contains contour plots for $\vert(Q/\omega^2)\vert
=10^{-2}$ at Hubble exit plotted as a function of $\beta$ and $k$.  
\begin{figure}[!htb]
\begin{center}
\epsfxsize 3.00 in
\epsfysize 2.50 in
\epsfbox{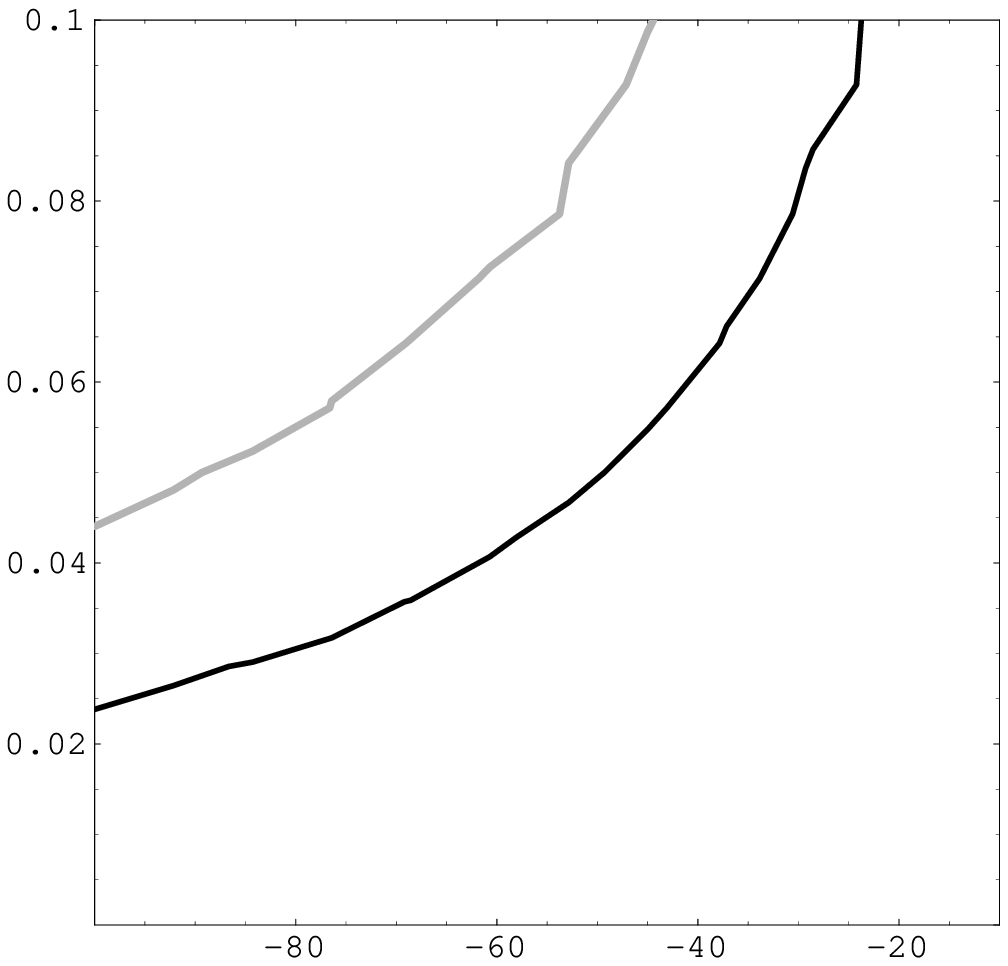}
\vskip -122 true pt \hskip -231 true pt \rotatebox{90}{$-(\beta+2)$}
\vskip 88 true pt \hskip 15 true pt {$\log_{10}(k/\Hbar)$}
\vskip 2 true pt
\caption{Contour plots for $\vert(Q/\omega^2)\vert=10^{-2}$ at 
Hubble exit plotted as a function of $\beta$ and $k$.  
The black contour corresponds to $(\Hbar/k_{c}) = 10^{-3}$ and 
the grey one to $(\Hbar/k_{c}) = 10^{-5}$. 
The WKB approximation is valid for values of $\beta$ and $k$ 
that lie below these contours.
In plotting these contours we have assumed that $\Hbar=10^{14}\, 
{\rm GeV}=10^{52}\, {\rm Mpc}^{-1}$.}
\label{fig:wkb}
\end{center}
\end{figure}
It is clear from the figure that the WKB approximation will be 
valid for smaller and smaller values of $k$, provided we choose 
correspondingly smaller and smaller values of $-(\beta+2)$. 
In the limit of exponential expansion, i.e. as $\beta\to -2$, the
WKB approximation, as we had shown earlier, turns out to be valid 
for all $k$. 

A few clarifying remarks concerning the above WKB approximation for
the massive modes are in order at this stage of our discussion.  Even
in the case of the standard massless and minimally coupled scalar
field, it is well-known that, while the WKB approximation is valid at
the sub-Hubble and at the super-Hubble scales, it breaks down at
Hubble exit (for a detailed discussion on this point and the
evaluation of the standard spectrum in the WKB approximation, see
Ref.~\cite{ms03}).  In fact, the WKB approximation will break down at
Hubble exit even for a massive field if the mass of the field is
smaller than (or of the order of) the Hubble scale $\Hbar$.  This is
evident from Eq.~(\ref{eq:WKBveiHe}) for the case of exponential
expansion.  However, in our model, we have assumed that the mass
(viz. $k_{\rm c}$) of the field to be very large compared to $\Hbar$.
It is due to this reason that we find the WKB approximation to be
valid at Hubble exit for all values of $k$ in the case of exponential
inflation and up to a certain minimum value of $k$, which depends on
the values of $\beta$ and $(\Hbar/k_{\rm c})$, in the case of power
law inflation.

The general WKB solution to the equation~(\ref{eq:demukcux}) can be 
written as
\br
\!\!\!\!\!\!\!\!\!\!
{\bar u}_{k}(x) 
&\simeq &\biggl[\l(\frac{{\bar A}(k)}{\sqrt{\omega(x)}}\r)\, 
\exp\, i\int^{x}dx'\,\omega(x')\nn\\ 
& &\qquad+\, \l(\frac{{\bar B}(k)}{\sqrt{\omega(x)}}\r)\, 
\exp\,-i\int^{x}dx'\,\omega(x')\biggr],\;\;
\er
where ${\bar A}(k)$ and ${\bar B}(k)$ are $k$-dependent constants that
are to be fixed by the initial conditions.  As mentioned earlier, we
shall assume that the massive field is in the Bunch-Davies vacuum on
the sub-Hubble scales.  (Note that, in obtaining the standard
spectrum, the massless field is assumed to be in the Bunch-Davies
vacuum.  Any possible excitations of the field by the evolving
background are expected to be suppressed exponentially by the mass
(see, for e.g., Ref.~\cite{bd82}).  Therefore, it is natural to assume
that the massive field---in particular, a field with a mass
(viz. $k_{\rm c}$) much greater than the Hubble scale---is in the
Bunch-Davies vacuum as well.)  Hence, the modes ${\bar u}_{k}$ are
required to have the limiting form~(\ref{eq:sticux}) which leads to
the conditions that ${\bar B}(k) = 0$ and ${\bar A}(k) =
\l[(\beta+1)/2 k\r]^{1/2}$.  Therefore, the mode ${\bar u}_{k}(x)$ is
given by
\beq
{\bar u}_{k}(x) 
\simeq \l(\frac{\beta+1}{2k\, \omega(x)}\r)^{1/2}\, 
\exp\, i\int^{x}d{x'}\, \omega(x')
\eeq
and, using these modes, it is straightforward to evaluate the 
modified perturbation spectrum~(\ref{eq:psm}). 
We find that the resulting spectrum can be written as
\br
\!\!\!\!\!\!\!\!\!\!\!\!\!\!
\left[k^3\; {\cal P}_{\Psi}(k)\right]_{\rm M}
&\simeq &C\,\l(\frac{\Hbar^2}{2\pi^2}\r)\,
\l(\frac{k}{\Hbar}\r)^{2 (\beta + 2)}\nn\\
& &\quad\,\times\l[1 - {\bar C}\,\l(\frac{\Hbar}{k_{c}}\r)\, 
\l(\frac{k}{\Hbar}\r)^{(\beta + 2)}\r],\label{eq:psmfr}
\er
where ${\bar C}$ is given by 
\beq
{\bar C} = \l[2\, C\, \l(\beta +1\r)^{3(\beta + 1)}\r]^{-1}.
\eeq

The modified spectrum~(\ref{eq:psmfr}) has some similarities to the
power spectrum that has been obtained recently from non-commutative
inflation~\cite{nci} and the spectrum exhibits a suppression of power
at the large scales.  In order to illustrate this feature, we have
plotted the modified spectrum~(\ref{eq:psmfr}) as well as the standard
spectrum, normalized to $\Hbar^2$, in Fig.~\ref{fig:comp} below.
(Note that, in plotting the modified spectrum, we have chosen the
values of $(\Hbar/k_{\rm c})$ and $\beta$ so that the WKB
approximation is valid for the massive modes over the range of $k$ of
interest.)
\begin{figure}[!htb]
\begin{center}
\epsfxsize 3.00 in
\epsfysize 2.30 in
\epsfbox{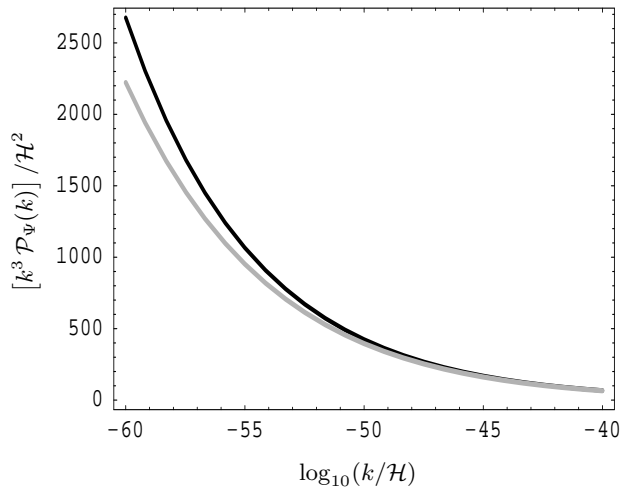}
\vskip -120 true pt \hskip -236 true pt \rotatebox{90}{$\l[k^3\, 
{\cal P}_{\Psi}(k)\r]/\Hbar^2$}
\vskip 60 true pt \hskip 15 true pt $\log_{10}(k/\Hbar)$
\vskip 2 true pt
\caption{Plots of the standard power spectrum $\l[k^3\; 
{\cal P}_{\Psi}(k)\r]$ (black curve) and the modified 
power spectrum  $\l[k^3\; {\cal P}_{\Psi}(k)\r]_{\rm M}$ 
(grey curve), normalized to $\Hbar^2$.
The grey curve corresponds to $(\Hbar/k_{\rm c})= 10^{-3}$ and 
$\beta = - 2.04$---values that have been chosen so that the WKB
approximation is valid around $k\sim 10^{-4}\, {\rm Mpc}^{-1}$.
In plotting these spectra we have assumed that $\Hbar=10^{14}\, 
{\rm GeV}=10^{52}\, {\rm Mpc}^{-1}$.}
\label{fig:comp}
\end{center}
\end{figure}
It is evident from the figure that the modified spectrum exhibits a
suppression of power around $\log_{10}(k/\Hbar)\sim -56$ which
corresponds to $k \sim 10^{-4}\, {\rm Mpc}^{-1}$ or $l =2, 4$---a
feature that seems to be necessary to explain lower power in the
quadrupole ($l = 2$) and the octopole moments ($ l = 4$) in the CMB
(see, for e.g., Ref.~\cite{blwe03}).  Though the modified spectrum we
have obtained (by using the standard inflationary parameters) shows a
suppression of power around the expected values of $k$, the extent of
the suppression proves to be far less than that is required by the
WMAP data.  In order to fit the WMAP data, one seems to require a
spectrum of the following form (see Ref.~\cite{contaldi}; in this
context, also see Ref.~\cite{tarun}):
\beq
\left[k^3\; {\cal P}_{\Psi}(k)\right]_{\rm M}
= A_{\rm s}\, k^{(n_{\rm s}-1)}\,
\biggl[1 - \exp-\l(k/k_\ast\r)^\gamma\biggr],\label{eq:psmwmap}
\eeq
where $A_{\rm s}$ and $n_{\rm s}$ are the amplitude and index of the
standard spectrum, $k_{\ast}\simeq 5\times 10^{-4}\, {\rm Mpc}^{-1}$ 
and $\gamma \simeq 3.35$.
To illustrate the fact that our model predicts far less suppression
than is required by the observations, it is useful to consider the
following ratio of the modified spectrum and the standard spectrum:
\beq
{\cal R}(k)
=\l(\frac{\left[k^3\; {\cal P}_{\Psi}(k)\right]_{\rm M}}
{\left[k^3\; {\cal P}_{\Psi}(k)\right]}\r).
\eeq
In Fig.~\ref{fig:ps} below, we have plotted the ratio ${\mathcal
R}(k)$ for the cases of the modified spectrum predicted by our model
[viz.~(\ref{eq:psmfr})] and the modified spectrum~(\ref{eq:psmwmap})
that seems to be required to fit the observations.  [Note that the
ratio ${\mathcal R}(k)$ for these spectra is given by the expressions
within the square brackets in Eqs.~(\ref{eq:psmfr})
and~(\ref{eq:psmwmap}).]
\begin{figure}[!htb]
\begin{center}
\epsfxsize 3.00 in
\epsfysize 2.30 in
\epsfbox{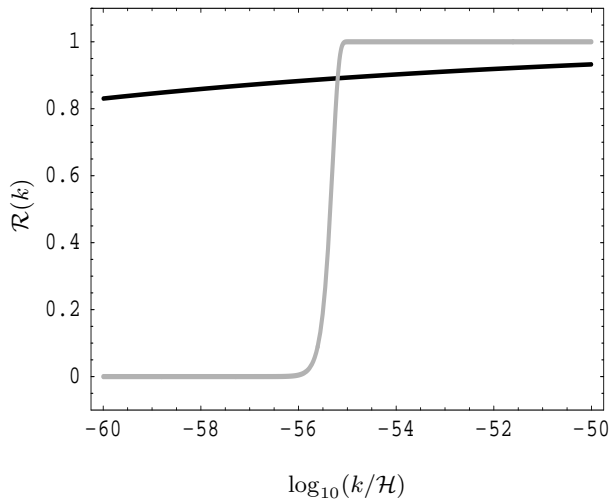}
\vskip -102 true pt \hskip -231 true pt \rotatebox{90}{${\cal R}(k)$}
\vskip 88 true pt \hskip 15 true pt $\log_{10}(k/\Hbar)$
\vskip 2 true pt
\caption{The ratio ${\cal R}(k)$ of the modified to the standard
spectrum plotted for the modified spectrum predicted by our 
model [viz.~(\ref{eq:psmfr})] and the spectrum~(\ref{eq:psmwmap})
that seems to be required to fit the WMAP data.
The black curve corresponds to the modified spectrum~(\ref{eq:psmfr})
for $(\Hbar/k_{c}) = 10^{-3}$ and $\beta=-2.04$---values that have been 
chosen so that the WKB approximation holds good around 
$k\sim 10^{-4}\, {\rm Mpc}^{-1}$.
The grey curves corresponds to the spectrum~(\ref{eq:psmwmap})
for $k_{\ast}= 5\times 10^{-4}\, {\rm Mpc}^{-1}$ and $\gamma=3.35$. 
Note that, in plotting these curves we have assumed that $\Hbar=10^{14}\, 
{\rm GeV}=10^{52}\, {\rm Mpc}^{-1}$.}
\label{fig:ps}
\end{center}
\end{figure}
It is evident from the figure that, in power law inflation, the 
suppression of power predicted by our model for the scalar power 
spectrum is far less than that is required to fit the WMAP data.

We would like to stress here the following points regarding the
modified spectrum we have obtained.  Firstly, one would tend to assume
that the high energy effects will leave their imprints only at the
ultra-violet end of the primordial perturbation spectrum.  In
contrast, we find that, what are supposedly trans-Planckian effects,
result in a modification of the spectrum at the infra-red end.  This
essentially arises due to the fact that the longer wavelength modes
leave the Hubble radius at earlier epochs thereby carrying the
imprints of the high energy effects\footnote{We thank Robert
Brandenberger for drawing our attention to this point.}.  Secondly,
assuming that the Planck scale effects can be expected to improve the
divergence structure of quantum field theory, we had expressed the
modified Wightman function~(\ref{eq:mGfnx}) as the {\it difference}\/
of the massless and the massive Wightman functions.  Due to this
reason and, also since the correction term to the power spectrum due
to the massive modes is positive definite [cf.  Eq.~(\ref{eq:psm})],
it is clear that the corrections will always suppress the power rather
than enhance it.  However, evidently, the actual form and the extent
of the suppression will depend on the correction term.  Though,
interestingly enough, we find that the massive modes indeed suppress
the power at large scales as required by the observations, the amount
of suppression proves to be {\it far less}\/ than that is needed to
fit the WMAP data.  Fourthly, we should point out that our model does
not contain any free parameters other than the ratio $(\Hbar/k_{\rm
c})$ and, hence, it can be said to predict the loss of power at large
scales naturally.  Finally, as we had mentioned earlier, we have
evaluated the trans-Planckian corrections to the standard, adiabatic
spectrum when the massive modes leave the Hubble radius.  In the
standard inflationary picture, the amplitude of the spectrum of the
massive modes would decay---the correction term in the
spectrum~(\ref{eq:psmfr}) would retain its spectral shape, but its
amplitude would decay as $e^{3(\beta+1)x}$---at the super Hubble
scales (i.e. as $x\to\infty$).  Since, the massive modes in our model
represent trans-Planckian corrections to the standard massless modes,
we have assumed that the mechanism that `freezes' the amplitude of the
standard spectrum at super Hubble scales will also `freeze' the
amplitude of the corrections at their value at Hubble exit.

\section{Discussion}

In this work, we have studied the trans-Planckian effects on the
spectrum of the primordial density perturbations in the power-law
inflationary scenario using an approach that preserves local Lorentz
invariance.  Motivated by the fact that quantum gravitational effects
can be expected to improve the divergence structure of standard
quantum field theory, we {\it assumed}\/ that the trans-Planckian
physics modifies the standard massless propagator such that the
modified propagator can be expressed as the {\it difference}\/ of the
original massless propagator and a massive propagator with a mass of
the order of the Planck mass.

In the standard inflationary scenario, the primordial perturbation
spectrum is determined by the Fourier transform of the massless scalar
field propagator.  Therefore, modifying the scalar field propagator
leads to modifications in the perturbation spectrum.  We find that, in
our model, the resulting modified spectrum remains scale invariant at
the ultra-violet end, but, interestingly, it exhibits a suppression of
power at the infra-red end---a feature that seems to be necessary to
explain the low quadrupole and octopole moments measured in the
CMB~\cite{MAP,contaldi,blwe03,tarun}.  However, at Hubble exit, the
amount of suppression predicted by our model in power-law inflation
turns out to be {\it far less}\/ than as expected from WMAP data.
Nevertheless, the loss of power at small $k$ suggests that the power
spectrum we have obtained may fit the WMAP data better than the
standard $\Lambda$CDM model.  It will be interesting to analyze the
implications of the WMAP data for our model in the context of
slow-roll inflation~\cite{ss04}.

Naively, one would expect that very high energy effects will leave
their imprints only at the ultra-violet end of the primordial
spectrum.  However, we find that the high energy effects lead to a
modification of the spectrum at the infra-red end.  As we had pointed
out in the last section, this can be attributed to the fact that the
longer wavelength modes leave the Hubble radius at earlier epochs
thereby carrying the signatures of the high energy effects.

Finally, we would like to emphasize a very attractive feature of our
approach which can result in corrections at the infra-red {\it and}\/
the ultra-violet ends of the perturbation spectrum.  Recall that in
obtaining the modified spectrum (\ref{eq:psmfr}) we had assumed that
both the massless and the massive fields are in the Bunch-Davies
vacuum.  This need not be the case.  While it is natural to assume
that the massive field is in the Bunch-Davies vacuum (in particular, a
very heavy field with a mass of the order of the Planck mass), one can
assume the massless field to be in a state such as the minimal
uncertainty state~\cite{minimaltp}.  In such a case, the resulting
modified spectrum will have corrections at both the ends.  While at
the infra-red end, we will simply reproduce the corrections we have
obtained, in the ultra-violet limit, the corrections will be of the
form that has been recently obtained in the literature~\cite{act03}.

\section*{Acknowledgments}

The authors wish to thank T.~Padmanabhan and Robert Brandenberger for
their comments on {an earlier version of the manuscript}.  LS wishes
to thank Namit Mahajan for discussions.  LS would also like to thank
the High Energy Physics Group of The Abdus Salam International Centre
for Theoretical Physics, Trieste, Italy for hospitality, where this
work was completed.

\end{document}